\documentclass[applsci,review,accept,moreauthors,10pt,a4paper]{mdpi}

\firstpage{1}
\articlenumber{362}
\doinum{10.3390/app8030362}
\pubvolume{8}
\pubyear{2018}
\copyrightyear{2018}
\history{Received: 26 January 2018; Accepted: 18 February 2018; Published: 2 March 2018}
\usepackage{siunitx}

\usepackage{color}

\setitemize{parsep=6pt,itemsep=0pt,leftmargin=*,labelsep=5.5mm}
\setenumerate{parsep=6pt,itemsep=0pt,leftmargin=*,labelsep=5.5mm}
\setlist[description]{itemsep=0mm}

\Title {Orbital Angular Momentum Generation and Detection by Geometric-Phase Based Metasurfaces}

\Author{Menglin L. N. Chen $^{1,2}$, Li Jun Jiang $^{1,2,}$* and Wei E. I. Sha $^{3,}$*}

\AuthorNames{Firstname Lastname, Firstname Lastname and Firstname Lastname}

\address{%
$^{1}$ \quad Department of Electrical and Electronic Engineering, The~University of Hong Kong, Hong Kong, China; menglin@connect.hku.hk\\
$^{2}$ \quad HKU Shenzhen Institute of Research and Innovation, Shenzhen 518057, China\\
$^{3}$ \quad Key Laboratory of Micro-Nano Electronic Devices and Smart Systems of Zhejiang Province, \mbox{College of Information Science and Electronic Engineering}, Zhejiang University, Hangzhou 310027, China}

\corres{Correspondence: jianglj@hku.hk (L.J.J.); weisha@zju.edu.cn (W.E.I.S.)}

\abstract{We present a comprehensive review on the geometric-phase based metasurfaces for orbital angular momentum (OAM) generation and detection. These metasurfaces manipulate the electromagnetic (EM) wave by introducing abrupt phase change, which is strongly dependent on the polarization state of incident EM wave and can be interpreted by geometric phase. Hence, the~conventional bulk devices that based on the accumulated phase change along the optical path can be avoided.}


\begin{document}


\section{Introduction}

It is well known that electromagnetic (EM) wave carries linear momentum which is associated with the poynting vector. Besides the linear momentum, EM wave has been demonstrated to possess angular momentum (AM), including spin angular momentum (SAM) and orbital angular momentum (OAM)~\cite{bliokh2013dual}. Circularly polarized wave carries SAM. The~SAM is $\hbar$ per photon for left-handed circular polarization (LHCP) and $-\hbar$ per photon for right-handed circular polarization (RHCP), where $\hbar$ is the reduced Planck constant. The~SAM characterizes the spin feature of photon. Different from the SAM, OAM manifests the orbital rotation of photon and each photon has an OAM of $l\hbar$, where $l$ is known as the OAM index and can be any integer. Different values of $l$ correspond to mutually orthogonal OAM states and the number of allowable OAM states for photons is unbounded. Different OAM states have been used to encode information in communications to enhance the channel capacity, both in free space~\cite{padgett2004free,willner2016experimental,willner2015optical,wangjian2012terabit,thide2007utilization,walker20134,yanyan2014high,zhangxianmin2015multiplexed} and optical fibers~\cite{willner2013terabit}. The~main problem of practical application of OAM in communications is the significant crosstalk between OAM modes~\cite{bjork2015OAM,willner2015performance}. The spatial-dependence and divergency nature of OAM-carrying waves result in the their vulnerability to the atmosphere~\cite{paterson2005atmospheric,zheng2012aberration} and limited power received at the receiver side~\cite{thide2015space}. All these facts will degrade the purity of OAM modes. Still, OAM offers a new attractive degree of freedom to EM waves, extending beyond the existing wave features. The~optical vortices of OAM beams also find their applications in super resolution imaging~\cite{tamburini2006overcoming,liukang2015oam}. OAM can be transferred to particles, which has been applied in optical tweezers~\cite{grier2003a,dunpop1995direct,padgett1997mechanical}. The~exchange of OAM with matter can also be utilized in detecting the rotation of particles~\cite{padgett2013detection}. In~quantum mechanics, at a single-photon level, OAM modes can be utilized for high-dimensional entanglement~\cite{padgett1996experimental,padgett2010quantum}.  Overall, OAM has received great attentions in multidisciplinary research areas~\cite{gumin2016onchip,qiuchengwei2016onchip,torner2007twisted,zeilinger2002superpositions}.

Since OAM has great potentials in various applications, research has been undertaken extensively on its generation. In~1992, Allen {et al.} first found that a Laguerre-Gaussian (LG) beam with helical wavefront carries well defined OAM~\cite{allen1992oam}. That beam possesses an azimuthal phase term, $e^{il\phi}$, where $\phi$ is the azimuthal angle. Before that, Soskin {et al.} created structured light with helical wavefront by forked gratings~\cite{soskin1990laser}. This~grating forms the basis of the computer generated holograms (CGHs) for producing light beams with OAM. Afterwards, many other devices for OAM generation have been proposed, such as the cylindrical lens in 1993~\cite{woerdman1993astigmatic}, spiral phase plates (SPPs) in 1994~\cite{woerdman1994helical}, q~plates in 2006~\cite{paparo2006optical}. Lately, OAM has been analyzed at radio frequencies. Antenna arrays~\cite{thide2010OAM}, traveling-wave antennas~\cite{zhangxianmin2015transmission} and circularly polarized patch antennas~\cite{toscano2014circular} which radiate EM waves carrying OAM have been demonstrated.

Recent advances in versatile metasurfaces also expedite powerful and convenient design routes for OAM generation~\cite{capasso2014flat,capasso2011light,capasso2012ultrathin,capasso2013flat}. The~concept of metasurface originates from the conventional frequency selective surfaces (FSSs)~\cite{munk2000fss}. They are composed of man-made subwavelength scatterers with varying geometry and orientation. Therefore, they go beyond the conventional FSSs due to the high feasibility of tailoring the geometry and orientation of the ultracompact scatterers. Metasurfaces locally alter the wave properties by the abrupt phase change at the scatterers.  By varying the geometry or orientation, scatterers can cover a total $2\pi$ phase shift so that arbitrary beam forming can be achieved. Meanwhile, scatterers can be designed to simultaneously change the wave amplitude~\cite{lilong2017generation}. Besides the electric response, they can have the magnetic response. Scatterers have both electric and magnetic responses form the Huygens metasurface~\cite{grbic2013metamaterial}. The~magnetic response helps to compensate the impedance mismatch at the metasurface interface so that nearly perfect efficiency can be achieved.

Although there are various designs in different types and suitable for different application scenarios, the~design principles for OAM generation lead to one common rule: the introduction of the azimuthal phase term $e^{il\phi}$ to EM waves. Generally, they fall into two categories: independent and dependent on the wave polarization.

The~first scheme employs isotropic materials, such as SPPs, CGHs. $e^{il\phi}$ is introduced in a SPP based on the accumulated spatially varying optical path~\cite{hasman2005spiral,padgett1996generation,soifer2005generation}. They are usually implemented at optical frequencies. At lower frequencies, the~device would become bulky unless some flatten techniques are employed~\cite{zhangianmin2015ultralow,hongwei2014generation,litchinitser2014spinning}. CGHs are diffractive optical elements and they produce light with different OAMs according to the diffraction orders~\cite{salgueiro2008making,white1992generation,padgett1998the,cottrell2009vortex}. Alternatively, the~excitation of an antenna or antenna array can be modulated directly to satisfy the required phase condition so that an OAM wave can be radiated out~\cite{liukang2016generation,ettorre2017exciting}. $e^{il\phi}$ can also be produced based on the abrupt phase shift at scatterers on a metasurface. By varying the geometry of the scatterer, its resonant frequency is changed so that the phase shift varies at the designed frequency. A total $2\pi$ phase shift is achieved after optimization; and successful generation of different OAM states has been reported in~\cite{lihongqiang2016generating,lilong2016generating}. {{Lately, tunable scatterers loaded by varactor diodes are proposed for convenient multiple OAM-mode generation~\cite{shihongyu2017generation}.}} Scatterers can also be made anisotropic to achieve independent control of different polarizations. However, even if the response of scatterers is polarization dependent, the~helicity of the produced OAM does not depend on the polarization state of incident wave. In~other words, the~helicity is fixed. While for the second scheme, it is an opposite scenario.

The~second scheme is based on the conversion and coupling between SAM and OAM. This~process occurs in inhomogeneous and anisotropic media, such as q plates~\cite{gray2013q,santamato2012polarization}. Q plates shift the circular polarization state from left to right or right to left and have spatially varying optical axis~\cite{bliokh2015spin}. Their~behaviors can be explained by AM conservation law. The~flip of the circular polarization state indicates a change of $\pm2 \hbar$ in SAM. When the q plate is cylindrically symmetric, the~total AM is conserved so that the output wave must carry an OAM of $\mp2 \hbar$. When the q plate is not cylindrically symmetric, it introduces extra AM to the system so that different orders of OAM can be generated in the output wave. {{Q plates are usually made by liquid crystals, owning to their flexibility and anisotropic properties. By tuning the temperature or external voltage, the~birefringent retardation in liquid crystals varies, so that the conversion efficiency can be tuned~\cite{santamato2009efficient,santamato2010photon}. The~maximum efficiency is obtained when the retardation is $\pi$.}} Metasurfaces have been used to implement the feature of q plates~\cite{kangming2012wave,yuanfang2017broadband,boyd2014generating,zhuangshuang2017controlling,zhangshuang2013spin}. The~flip of the circular polarization state achieved by anisotropic scatterers and rotation of the scatterers is an analogue to the rotation of optical axis of a q plate, along with the introduction of geometric phase. Therefore, this type of metasurfaces is known as geometric-phase metasurfaces. Unlike the metasurfaces employing the first scheme, geometric-phase metasurfaces produce OAM along with the change of the polarization state of the output wave, which is fundamentally different from the first scheme. The~handness of the produced OAM depends on the incident SAM. Hence, geometric-phase metasurfaces show their high flexibility in the manipulation of OAM because they enable the coupling and interchange between SAM and OAM.

In~communications, detection of OAM at the receiver side is required. Although OAM detection is just a reciprocal process of OAM generation, it is much more challenging due to the degraded OAM states after the propagation. OAM detection can be classified into three categories, mode analysis based on field data~\cite{thide2010OAMmeasure,forbes2013measurement,zhangxianmin2016local}, observation of OAM induced effects such as the rotational Doppler shift~\cite{malu2017detecting,padgett1998measurement,allen1994azimuthal} and the beam reforming by adopting holographic technology~\cite{capasso2012holographic}. The~first approach needs to acquire all the three components of electric and magnetic fields so that the OAM  can be calculated explicitly~\cite{thide2010OAMmeasure}. Alternatively, one can measure the phase gradient that is equal to the OAM index $l$~\cite{thide2012encoding}. In~the third approach, by using a hologram, the~OAM modes can be projected into a detectable Gaussian mode~\cite{mair2001entanglement}. Geometric-phase metasurfaces can be used to implement the holograms and have the advantages of low profile and high tuning flexibility.

In~this review article, we concentrate on the current research of the OAM generation and detection by geometric-phase based metasurfaces. We start by the introduction of geometric phase. Then, we go over several novel metasurfaces for OAM generation. The~induced geometric phases on the metasurfaces can be in both discrete and continuous formats. Their~geometries, working principles and novel functionalities have been reviewed and discussed in detail. At last, a method for OAM detection is reviewed.

\section{Geometric Phase}

When a light changes its initial polarization state to a final polarization state along different paths on a Poincar\'e sphere, the~final polarization states will have difference phases, known as geometric phase~\cite{zhoulei2015photonic}. In~the following, we derive the geometric phase when the SAM shifts from $\pm \hbar$ to $\mp \hbar$ through an anisotropic scatterer with different orientations.

The~behavior of a scatterer can be modelled by Jones matrix $\mathbf{J}$. It connects the polarization state of the scattered wave with that of an incident wave:
\begin{equation}
\begin{pmatrix}
  E^s_x \\ E^s_y
\end{pmatrix}
=
\begin{pmatrix}
    J_{xx} & J_{xy} \\
    J_{yx} & J_{yy}
  \end{pmatrix}
\begin{pmatrix}
  E^i_x \\ E^i_y
  \end{pmatrix}
  =
    \mathbf{J}^{l}
\begin{pmatrix}
  E^i_x \\ E^i_y
\end{pmatrix}
,
\label{eq:1}
\end{equation}
where $E^i_x$ and $E^i_y$ are the $x$ and $y$ components of the incident electric field. $E^s_x$ and $E^s_y$ are the corresponding components of the scattered electric field.

Jones matrix under the circular basis can be obtained by coordinate transformation,
\begin{equation}
\mathbf{J}^{c}=
\begin{pmatrix}
    J_{++} & J_{+-} \\
    J_{-+} & J_{--}
  \end{pmatrix}
=
  \frac{1}{2}
  \begin{pmatrix}
    (J_{xx}+J_{yy})+i(J_{xy}-J_{yx}) & (J_{xx}-J_{yy})-i(J_{xy}+J_{yx})\\
    (J_{xx}-J_{yy})+i(J_{xy}+J_{yx}) & (J_{xx}+J_{yy})-i(J_{xy}-J_{yx})
  \end{pmatrix}
  ,
\label{eq:2}
\end{equation}
where $+$ and $-$ represent the right circularly polarized (RCP) and left circularly polarized (LCP) components.

If the scatterer achieves a perfect conversion between RCP wave and LCP wave, it must satisfy $J_{yy}=-J_{xx}$ and $J_{yx}=J_{xy}=0$, so that the Jones matrix is written as,
\begin{equation}
\mathbf{J}_{pol}^{l}
=
\begin{pmatrix}
J_{xx} & 0 \\
0 & -J_{xx}
\end{pmatrix}, \quad
\mathbf{J}_{pol}^{c}=
\begin{pmatrix}
0 & J_{xx} \\
J_{xx} & 0
\end{pmatrix}
\label{eq:3}
\end{equation}

Clearly, there are only off-diagonal entities in $\mathbf{J}^{c}_{pol}$, indicating the flip between RHCP and LHCP. Axially rotating the scatterer by an angle of $\alpha$ will result in a new Jones matrix,
\begin{equation}
\mathbf{J}^{l}_{pol}(\alpha)
=J_{xx}
\begin{pmatrix}
\cos(2\alpha) & \sin(2\alpha) \\
\sin(2\alpha) & -\cos(2\alpha)
\end{pmatrix}, \quad
\mathbf{J}^{c}_{pol}(\alpha)
=J_{xx}
\begin{pmatrix}
0 & e^{-2i\alpha} \\ e^{2i\alpha} & 0
\end{pmatrix}
\label{Eq:4}
\end{equation}

$\mathbf{J}^{c}_{pol}(\alpha)$ also has only the non-zero off-diagonal items. Besides, an additional phase factor $e^{\pm 2i\alpha}$ is introduced. This~phase is the so-called geometric phase. The scatterers on a geometric-phase metasurface cover a $2\pi$ phase shift by changing the rotation angle $\alpha$.

To generate OAM, geometric phase is utilized to construct the required phase profile, $e^{il\phi}$ of an OAM wave. Apparently, to generate an OAM of order $l$, scatterers on a metasurface with azimuthal location $\phi$ should be designed to provide a phase change of $l\phi$, i.e., $\alpha=\pm l\phi/2$. The~sign depends on the incident circular polarization state. The~value of $\alpha/\phi$ is known as the topology charge $q$ of the metasurface. Apparently, the~OAM order $l$ is determined by $q$ and should be double of its value.

The~geometric phase itself does not depend on the frequency so that broadband metasurfaces could be potentially designed. For~a perfect (100\%) conversion, the~condition $J_{yy}=-J_{xx}=\pm1$ and $J_{yx}=J_{xy}=0$ should be strictly satisfied, i.e., $J_{yy}$ and $J_{xx}$ should have the same unit amplitude and $\pi$ phase difference. Achieving these conditions by scatterers on resonance will inevitably limit the bandwidth. When the requirement is not strictly satisfied, the~co-circularly polarized component needs to be filtered out. Only the cross-circularly polarized component carries OAM.

\section{Metasurfaces for OAM Generation}

Various prototypes of scatterers that can totally or partially convert the R/LHCP to L/RHCP have been proposed for OAM generation. Plasmonic scatterers, such as the golden nanorods in~\cite{zhangshuang2017rotational}, L-shaped gold nanoantennas~\cite{boyd2014optical}, rectangular split-ring resonators (SRRs) in~\cite{qushiliang2015ultra}; apertures opened on metals, such as the elliptical nanoholes in~\cite{luoxiangang2016generation}; and dielectric particles have been demonstrated at optical regime. At microwave frequencies, multi-layered unit cells have been reported to achieve high efficiency~\cite{fusco2013split,lilianlin2015an}. {{The~scatterers and their composite metasurfaces are usually simulated by numerical simulation software, such as finite element method and finite integration technique-based CST Microwave Studio~\cite{menglin_pecpmc,menglin_csrr,luoxiangang2017OAM,luoxiangang2016merging,luoxiangang2015catenary,menglin_detect}, finite element method-based HFSS, finite-difference time-domain method-based Lumerical~\cite{capasso2017spin}, and COMSEL. Some of them can be analytically modelled~\cite{luoxiangang2016merging} or simulated by approximate models~\cite{menglin_pecpmc,menglin_detect}. Lately, an efficient modelling is applied to integral equation solvers to accelerate the simulation of metasurfaces in multiscale~\cite{limaokun2017quasi}.}}

\subsection{Artifical PEC-PMC Metasurface}

\begin{figure}[htbp]
\centering
\includegraphics[width=5.6in]{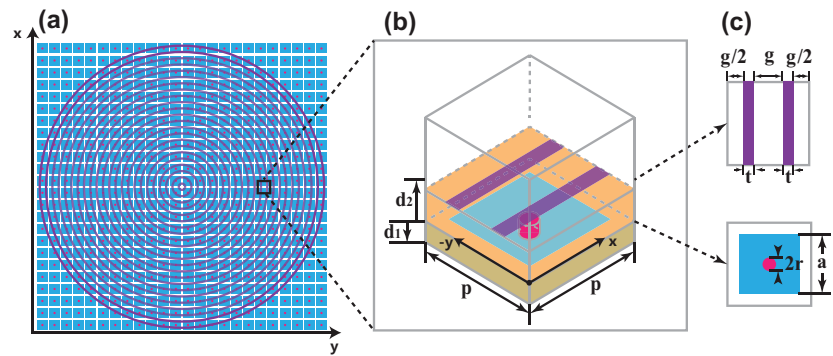}
\caption{Schematic pattern of the perfect electric conductor (PEC)-perfect magnetic conductor (PMC) anisotropic metasurface for orbital angular momentum (OAM) generation. With a nearly 100\% conversion efficiency, the~metasurface perfectly converts a left (right) circularly polarized plane wave carrying zero OAM to a right (left) circularly polarized vortex beam carrying $\pm 2\hbar$ OAM: (\textbf{a})~Top view of the whole metasurface. (\textbf{b},\textbf{c}) A scatterer in the metasurface. The~scatterer is composed of artificial PEC (purple) and PMC (blue and red) surfaces. The~period of the scatterer is $p=7$~mm. The~permittivity and thickness of the dielectric substrate are set to $\epsilon_{r}=2.2$, $d_1=2$~mm and $d_2=3$~mm. For~the artificial PEC surface (top-right inset), the~width and gap for the strip is $t=1$~mm and $g=2.5$~mm, respectively. For~the mushroom-based artificial PMC surface (bottom-right inset), the~square patch size is $a=6$~mm. A metallic via with the radius of $r=0.25$~mm and height of $d_1=2$~mm connects the patch to the ground plane. Reproduced with permission from~\cite{menglin_pecpmc}, Copyright AIP Publishing LLC, 2016.}
\label{Fig:unit_cell}
\end{figure}

We propose a composite perfect electric conductor (PEC)-perfect magnetic conductor (PMC)-based anisotropic metasurface~\cite{menglin_pecpmc}. A PEC surface is designed for $x$ polarization so we have $J_{xx}=-1$. An artificial PMC surface is put beneath the PEC surface so that $J_{yy}=1$ is satisfied. The~proposed novel metasurface is depicted in Figure~\ref{Fig:unit_cell}. The~metasurface has two dielectric layers and a ground plane. It is composed of scatterers shown in Figure~\ref{Fig:unit_cell}b. The~metal strips on the top layer function as a parallel plate waveguide and allow the perfect transmission of $y$ polarized wave while the $x$ polarized wave will be totally reflected with $\pi$ phase shift. The~metal patch in middle together with the via and ground plane forms the well-known mushroom-like high-impedance surface. By operating at its resonant frequency, it can be considered as a PMC surface and will reflect the transmitted $y$ polarized wave from the metal strips with a zero phase shift. The~whole metasurface is built by distributing the scatterers with varying orientations according to their azimuthal locations. It should be noted that the mushroom-like high-impedance surface is isotropic, therefore, there is no need to change the orientations of mushrooms. For~the PEC surface, by setting the rotation angle $\alpha=\phi$, rotation of the strips will lead to a series of concentric loops. The~resultant generated OAM order is $2$ or $-2$, where the sign depends on the incident circular polarization state.

\begin{figure}[htbp]
\centering
\includegraphics[width=5.6in]{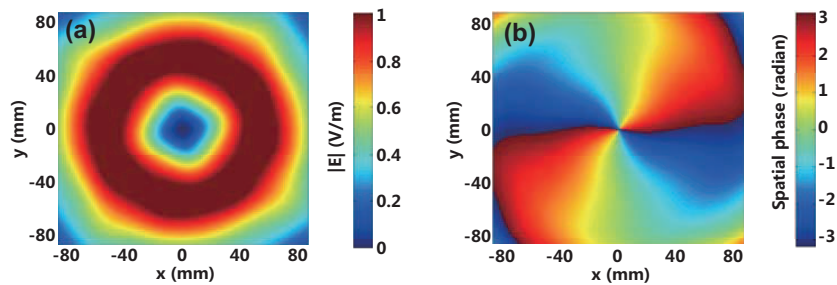}
\caption{The~amplitude and phase distributions of reflected electric fields from the PEC-PMC metasurface at a transverse plane $z=20$~mm: (\textbf{a}) Amplitude. (\textbf{b}) Phase. Reproduced with permission from~\cite{menglin_pecpmc}, Copyright AIP Publishing LLC, 2016.}
\label{Fig:pecmushroom}
\end{figure}

The~metasurface is designed at $6.2$ GHz {{and can be conveniently fabricated using printed circuit board (PCB) technique.}} Full-wave simulation was done in CST MWS. The~right circularly polarized plane wave is used as incident wave. Figure~\ref{Fig:pecmushroom} shows the amplitude and phase distributions of the reflected electric fields at a transverse plane of $z=20$~mm. An amplitude null can be observed in Figure~\ref{Fig:pecmushroom}a, which results from the phase singularity at the center for an OAM-carrying wave. In~Figure~\ref{Fig:pecmushroom}b, the~phase accumulated along a full circular path around the center is $4\pi$, indicating an OAM of order $-2$. A unique feature for the PEC-PMC metasurface is that the scatterers on top layer are not on resonance or discrete but continuously connected. Thus, the~near-field pattern is quite smooth without any evanescent field component scattered by discrete scatterers.

\begin{figure}[htbp]
\centering
\includegraphics[width=5.6in]{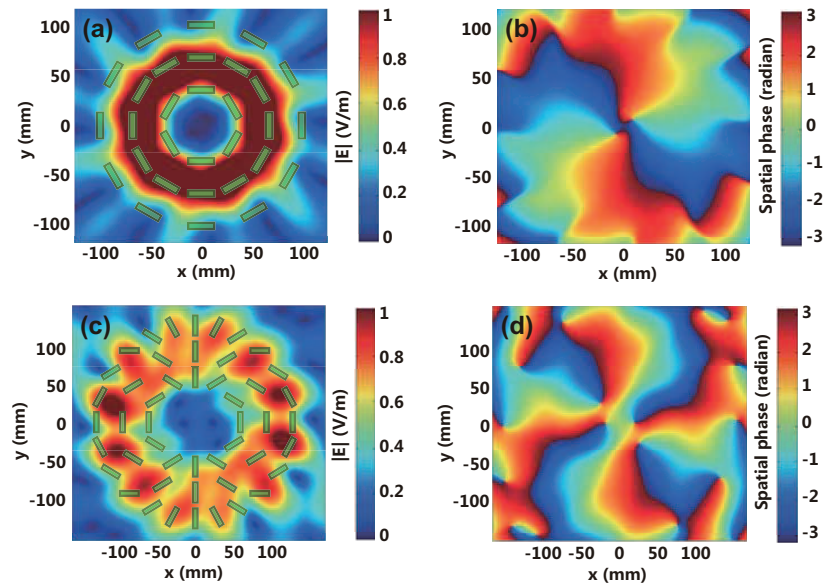}
\caption{The~amplitude and phase distributions of reflected electric fields from the discrete PEC-PMC metasurface. For~the generation of OAM of order $-2$, (\textbf{a}) amplitude; (\textbf{b}) phase at a transverse plane $z=40$~mm. For~the generation of OAM of order $-4$, (\textbf{a}) amplitude; (\textbf{b}) phase at a transverse plane $z=100$~mm. Reproduced with permission from~\cite{menglin_pecpmc}, Copyright AIP Publishing LLC, 2016.}
\label{Fig:dipole}
\end{figure}

For~the generation of OAM of other orders, a similar PEC-PMC metasurface but with discrete dipole scatterers on the top layer is proposed. The~dipoles are designed to totally reflect the $x$ polarized wave with $\pi$ phase shift at $6.2$~GHz. Two cases for generation of OAM of order $-2$ and $-4$ are shown in Figure~\ref{Fig:dipole}. The~amplitude and phase of the reflected field are drawn at a transverse plane of $z=40$~mm. The~distributed dipole scatterers are also depicted in Figure~\ref{Fig:dipole}a,c. The~field patterns verify the desired OAMs have been generated. However, due to the influence of the discrete dipoles, the~amplitude distribution is not uniform any more and ripples are observed in the phase distribution. As discussed before, the~ripples are caused by evanescent field components scattered by the dipoles. The~distortion only exists in the near field and will become less serious when the observation plane moves further away from the metasurface.

\subsection{Ultrathin Complementary Metasurface}

The~efficiency of reflective metasurface could be very high by placing a reflector beneath it, while for transmissive metasurface, high efficiency cannot be easily achieved due to the reflection at the metasurface-air interface stemming from the impedance mismatch. In~the following, we show our designed double-layer complementary metasurface for OAM generation with high efficiency~\cite{menglin_csrr}.

\begin{figure}[htbp]
	\centering
	\includegraphics[width=5.6in]{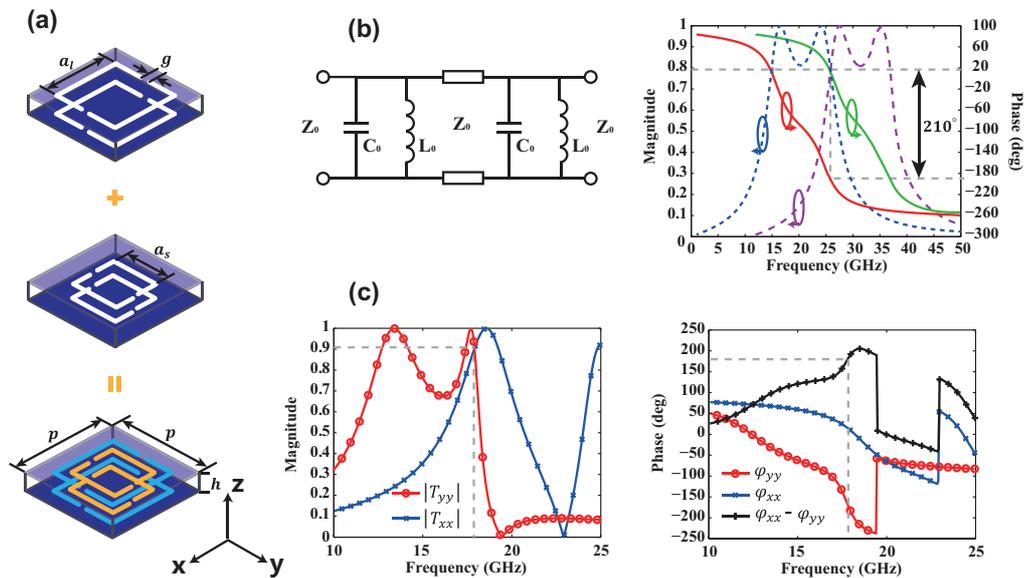}
	\caption{Schematic and the response of the proposed unit cell: (\textbf{a}) Schematic. (\textbf{b}) Equivalent circuit model of one pair of complementary split-ring resonators (CSRRs) and simulated $S_{21}$. The~purple and green curves are obtained by the translation of the original blue (magnitude) and red (phase) curves. The~distance between the two layers is $h$ = 1.25~mm. Characteristic impedance of free space is $Z_0=377~\Omega$. The~capacitance and inductance are $C_0=0.09$~pF and $L_0=1.03$~nH. (\textbf{c}) Full-wave simulation results. The~period of the unit cell is $7\times7$~mm$^2$. Side lengths of the two types of square CSRRs are $a_l=5.2$~mm and $a_s=3.9$~mm. The~length of the gap is $g=0.2$~mm. The~width of the slots is $t=0.2$~mm. Reproduced with permission from~\cite{menglin_csrr}, Copyright IEEE, 2017.}
	\label{unit}
\end{figure}

The~unit cell is shown in Figure~\ref{unit}a. Each unit cell consists of four complementary split-ring resonators (CSRRs) with two different sizes and orientations. The~equivalent circuit and its response for each pair are drawn in Figure~\ref{unit}b. It is clear that the double-layer structure offers two parallel inductor-capacitor (LC) resonant circuits so that we have a sufficient tunable range to achieve the design objective, i.e., the same high transmittance from the two pairs as well as the $\pi$ phase difference between their transmission coefficients. Figure~\ref{unit}c presents the full-wave simulation results of unit cell. At $17.85$~GHz, magnitudes of the co-transmission coefficients are both $0.91$ and their phase difference is $\pi$, indicating a $81\%$ circular polarization conversion efficiency. {{This~double-layer structure can be fabricated on a conventional PCB.}}

\begin{figure}[htbp]
	\centering
	\includegraphics[width=5.6in]{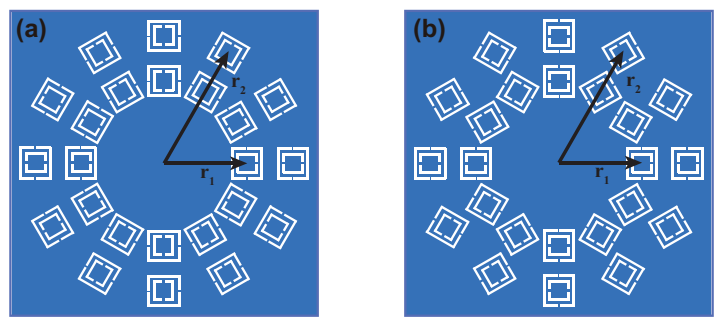}
	\caption{Geometric structure of the metasurface. Topological charge is (\textbf{a}) $q=1$; (\textbf{b}) $q=2$. The~radius of the inner ring is $r_1=14$~mm and that of the outer ring is $r_2=21$~mm. Reproduced with permission from~\cite{menglin_csrr}, Copyright IEEE, 2017.}
	\label{whole}
\end{figure}

Then a whole metasurface is built by arranging the unit cells with varying orientations. Two~metasurfaces with topological charges of 1 and 2 are designed. The~top view of the designed metasurface is shown in Figure~\ref{whole}. Each metasurface includes $24$ scatterers, whose centers describe two circles with the radius of $r_1$ and $r_2$.

\begin{figure}[htbp]
	\centering
	\includegraphics[width=5.6in]{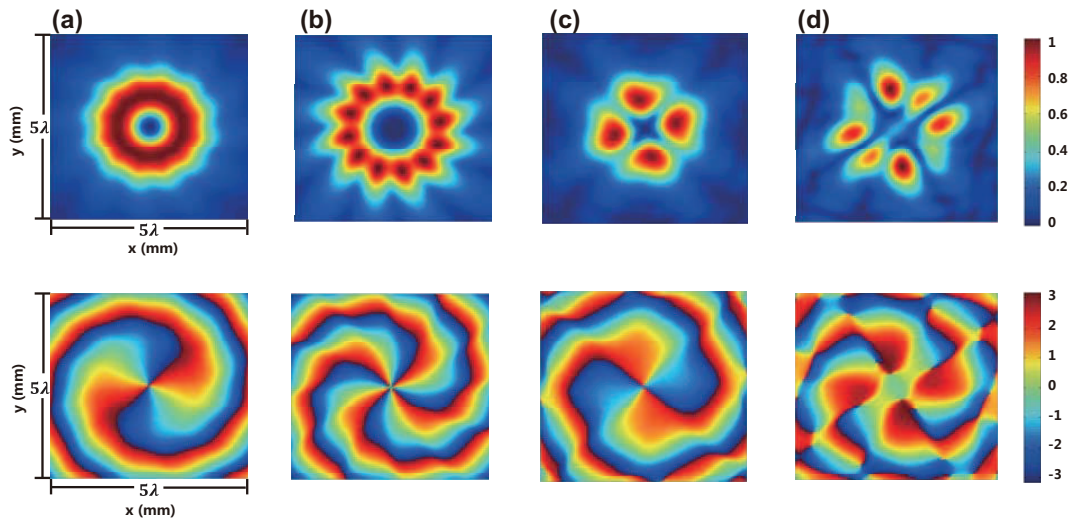}
	\caption{Amplitude and phase distributions of the cross-circularly polarized component of electric field at a transverse plane of $z=10$~mm calculated from (\textbf{a}) the equivalent dipole model with the aperture in Figure~\ref{whole}a, (\textbf{b}) the equivalent dipole model with the aperture in Figure~\ref{whole}b, (\textbf{c}) the full-wave simulation with the aperture in Figure~\ref{whole}a, (\textbf{d}) the full-wave simulation with the aperture in Figure~\ref{whole}b. Reproduced with permission from~\cite{menglin_csrr}, Copyright IEEE, 2017.}
	\label{field}
\end{figure}

The~EM responses from the metasurfaces are calculated by an equivalent dipole model and CST MWS. In~the equivalent dipole model, each unit cell is considered to be two orthogonal magnetic dipole sources with the orientations aligned with the long side of the CSRRs. Therefore, the~EM response from the whole metasurface is considered to be a sum of the response from magnetic dipoles with varying orientations. This~approach is useful to simulate a metasurface comprising a great number of scatterers that can be treated as point sources with arbitrary strengths, locations and orientations. The~calculated field distributions are shown in Figure~\ref{field}. The~results verify the successful generation of OAM. The~divergence between the results from the dipole model and the full-wave simulation comes from the coupling between unit cells.

\subsection{Metasurface Fork Gratings}

The~transmission function for a diffraction grating for OAM generation is written by
\begin{equation}
t(r,\phi) = \sum_{m} A_m e^{j(l_m\phi+k_{xm}x+k_{ym}y)},
\label{eq:t}
\end{equation}
where $r$ is the radial position, $\phi$ is the azimuthal position, $A_m$ is the weight of the $m$th beam, $l_m$ is the corresponding OAM index, and $k_{xm}$, $k_{ym}$ are the transverse wave numbers of the $m$th beam. It can be considered as a hologram resulting from the interference of a plane wave and a OAM wave.

\begin{figure}[htbp]
	\centering
	\includegraphics[width=5.6in]{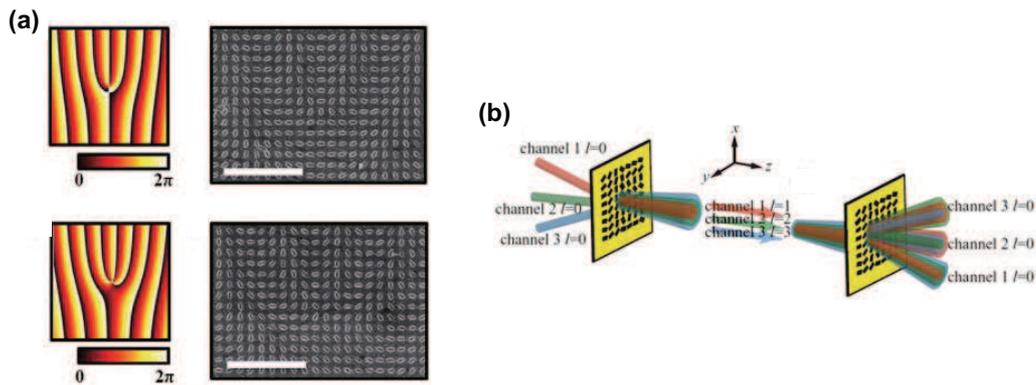}
	\caption{(\textbf{a}) Phase distribution and the scanning electron images of metasurface fork gratings with topological charge of $q = 2, 3$; The~plasmonic metasurfaces are fabricated on an $80$-nm thick aluminum thin film by using focus ion beam method, consists of spatially variant nanoslits with a size of $\sim$$50$~nm by $210$~nm. Scale bar: $3$~\SI{}{\micro\metre}. Reproduced with permission from~\cite{cheah2016geometric}, Copyright WILEY-VCH Verlag GmbH \& Co. KGaA, Weinheim, 2016. (\textbf{b}) Schematics of off-axis incidence multi-OAM multiplexer and off-axis multi-OAM demultiplexer. Reproduced with permission from~\cite{luoxiangang2017OAM}, Copyright WILEY-VCH Verlag GmbH \& Co. KGaA, Weinheim, 2016.}
	\label{cheah}
\end{figure}

The~transmission function is calculated based on the design requirement and the phase information is extracted and reconstructed using geometric-phase metasurfaces. We show two designs in Figure~\ref{cheah}~\cite{cheah2016geometric,luoxiangang2017OAM}. They are formed by nanoslits with varying orientations. Figure~\ref{cheah}a shows the phased holograms and the corresponding distributions of nanoslits for a single OAM beam generation. During the fabrication process, first, an 80-nm-think aluminum film is deposited on glass through thermal evaporation process. Then, the~nanoslits array is fabricated through focus-ion-beam method. Figure~\ref{cheah}b illustrates a general map of three-OAM-channel multiplexing and demultiplexing using metasurfaces. Two identical metasurfaces are used for the multiple OAM-beam generation and separation, respectively. In~both cases of Figure~\ref{cheah}, the~metasurfaces are illuminated by a circularly polarized wave and the cross-circularly polarized component carries OAM.

\subsection{Metasurface for OAM-Carrying Vector Beams Generation}

\begin{figure}[htbp]
	\centering
	\includegraphics[width=5.9in]{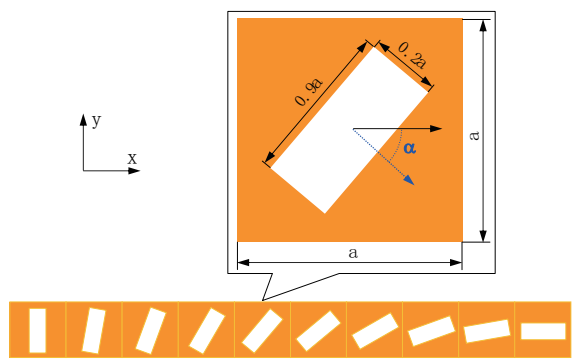}
	\caption{Geometry of a one-dimensional inhomogeneous anisotropic metamaterial composed of 10~rectangular holes with the orientation changed stepwisely from 0 to $\pi/2$. The~inset is the geometry of the unit cell that is a square metal slab punched into a rectangular hole. Reproduced with permission from~\cite{kangming2012twisted}, Copyright The~Optical Society, 2012.}
	\label{vector1}
\end{figure}

In~the previous discussion, the~metasurfaces generate circularly polarized waves with OAM. In~the following, we will review several pieces of research work dealing with vector fields carrying OAM~\cite{kangming2012twisted,willner2013metamaterials}. The~polarization of vector fields is represent by $\alpha(\phi)=m\phi+\alpha_0$, where $m$ is the polarization order and $\alpha_0$ is the initial polarization. To generate a vector field, a linear polarizer shown in Figure~\ref{vector1} is used. It composes of rectangular apertures and the transmitted wave is linearly polarized along the direction that is vertical to the long axis of the aperture. When the linear polarizers are oriented to different directions according to their locations, the~polarizations are varied locally. The~aperture with $\alpha=0$ is modelled by Jones matrix:
\begin{equation}
\mathbf{J}^{l}_{lin\_pol}
=
\begin{pmatrix}
1 & 0 \\
0 & 0
\end{pmatrix}.
\end{equation}

Then, the~Jones matrix for a rotated aperture is written as
\begin{equation}
\mathbf{J}^{l}_{lin\_pol}(\alpha)
=
\begin{pmatrix}
\cos^2(\alpha) & \sin(\alpha)\cos(\alpha) \\ \sin(\alpha)\cos(\alpha) & \sin^2(\alpha)
\end{pmatrix}
.
\label{Eq:6}
\end{equation}
\vspace{-12pt}

If the incident wave is circularly polarized, i.e., $E_{in}^{\pm}=\frac{1}{\sqrt{2}}[1 ~\pm i]^T$ (plus sign for LHCP and minus sign for RHCP), the~transmitted field is expressed by
\begin{equation}
E_{out}^+
=
\mathbf{J}^{l}_{lin\_pol}E_{in}^+
=
\frac{1}{\sqrt{2}}
e^{i\alpha}
\begin{pmatrix}
\cos(\alpha) \\
\sin(\alpha)
\end{pmatrix}
=
\frac{1}{2}
\begin{pmatrix}
1 \\
i
\end{pmatrix}
+
\frac{1}{2}
e^{2i\alpha}
\begin{pmatrix}
1 \\
-i
\end{pmatrix}
.
\label{Eq:7}
\end{equation}
\begin{equation}
E_{out}^-
=
\mathbf{J}^{l}_{lin\_pol}E_{in}^-
=
\frac{1}{\sqrt{2}}
e^{-i\alpha}
\begin{pmatrix}
\cos(\alpha) \\
\sin(\alpha)
\end{pmatrix}
=
\frac{1}{2}
e^{-2i\alpha}
\begin{pmatrix}
1 \\
i
\end{pmatrix}
+
\frac{1}{2}
\begin{pmatrix}
1 \\
-i
\end{pmatrix}
.
\label{Eq:8}
\end{equation}

From~\eqref{Eq:7} and~\eqref{Eq:8}, we can notice that the output wave is linearly polarized with a geometric phase of $e^{\pm i \alpha}$.  The~linearly polarized wave can be decomposed into two circularly polarized waves. The~co-circular polarization does not have the phase term and the cross-circular polarization has the term of $e^{\pm 2i \alpha}$. It should be emphasized that unlike the previous derivation that the conversion efficiency can reach 1, the~transmitted power after the linear polarizer is only half of the power in the incident wave and each transmitted circularly polarized component takes half of the total transmitted~power.

\begin{figure}[htbp]
	\centering
	\includegraphics[width=5.6in]{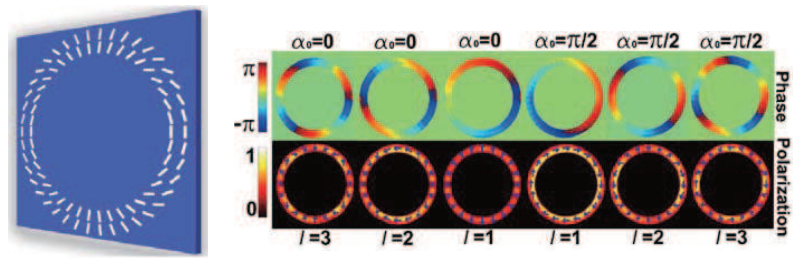}
	\caption{Schematic structure of metamaterials for generating OAM-carrying vector beams and the spatial distributions of phase and polarization of generated OAM-carrying vector beams ($\sigma=-1$, right circularly polarized (RCP)~input beam). Reproduced with permission from~\cite{willner2013metamaterials}, Copyright The~Optical Society, 2013.}
	\label{vector2}
\end{figure}

In~Figure~\ref{vector2}, two rings of slots are put on a gold film with thickness of $200$~nm. The~rotation angle of the slots $\alpha$ satisfies $\alpha(\phi)=l\phi+\alpha_0$. Thus, under the excitation of RCP light, according to Equation~\eqref{Eq:8}, the~spatially variant factor in $E_{out}$ is $e^{-i (l \phi + \alpha_0)}$. The~generated OAM order is $-l$. {{The~structure can be fabricated using electron-beam lithography~\cite{wangjian2016plasmonic}.}}

\subsection{Continuously Shaped Metasurfaces}

The~scatterers in the previous discussion provide discrete levels of abrupt phase shift by employing the geometric phase concept. In~this section, we review several prototypes with continuous or quasi-continuous phase levels~\cite{luoxiangang2016merging,luoxiangang2015catenary,hasman2002formation}. They generate OAM with high purity.

\begin{figure}[htbp]
	\centering
	\includegraphics[width=5.6in]{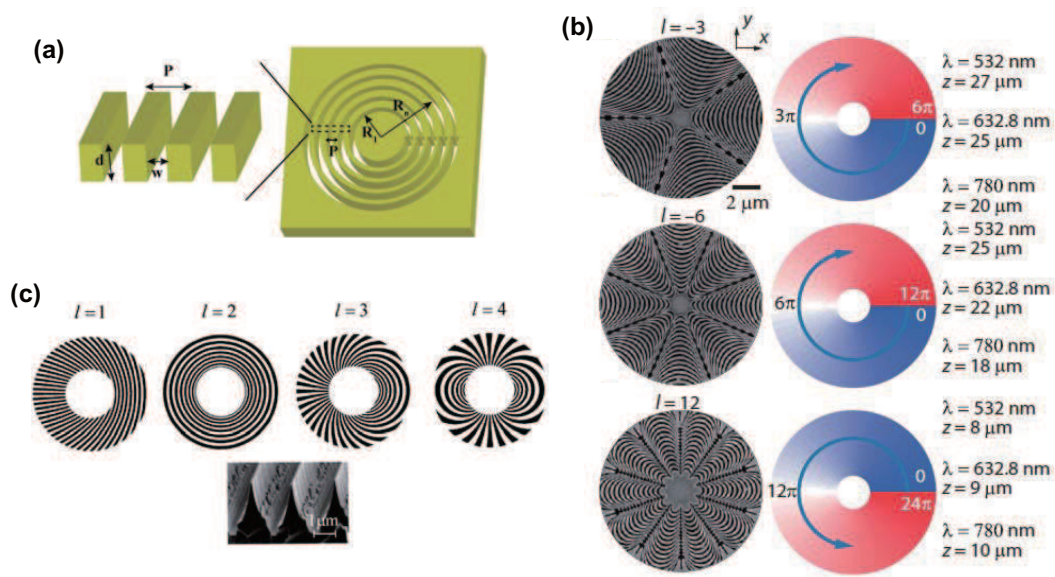}
	\caption{(\textbf{a}) The~metasurface is constructed by drilling a silver film with multiple periods of annular rings, whose radius is defined as $R_n=R_1+(n-1)P$, where $n$ and $P$ denote the number and the period of the apertures. The~annular apertures can be taken as two-dimensional extensions of a set of nanoslits with spatially varying orientation. Reproduced with permission from~\cite{luoxiangang2016merging}, Copyright American Chemical Society, 2016. (\textbf{b}) OAM generators based on catenary arrays. The~topological charges from up to bottom are $-$3, $-$6, and 12 ($s=1$), respectively. The~first column represents the scanning electron microscopy (SEM) images of the fabricated samples. The~second column shows the spiral phase profiles. Reproduced with permission from~\cite{luoxiangang2015catenary}, Copyright The~American Association for the Advancement of Science, 2015. (\textbf{c}) Top, geometry of the subwavelength gratings for four topological charges. Bottom, image of a typical grating profile taken with a scanning-electron microscope. Reproduced with permission from~\cite{hasman2002formation}, Copyright Optical Society of America, 2002.}
	\label{continuous}
\end{figure}

In~Figure~\ref{continuous}a, a metasurface is composed of annular apertures with smoothly changed widths. The~phase shift comes not only from the geometric phase due to the varying aperture orientation but also from the plasmon retardation phase which is modulated by the aperture width. With the contribution of the two phases, arbitrary OAM orders can be generated. Under the incidence of circularly polarized wave, according to~\eqref{Eq:8}, the~radially polarized component acquires a geometric phase of $e^{i\phi}$ and the cross-circularly polarized one has a geometric phase of $e^{2i\phi}$. {{To fabricate the metasurface sample, a 2-nm-thick Cr film and a 450-nm-thick silver film were subsequently deposited on a quartz substrate using magnetron sputtering. The~annular apertures are then milled on the silver/Cr film through focused-iron beam lithography.}} Figure~\ref{continuous}b shows the a pattern constructed from catenary-shaped atoms. The~inclination angle for a single caternary gradually varies a total of $pi$ from one end of the caternary to the other end. The~induced geometric phase doubles the value of inclination value for the cross-circularly polarized output wave. By arranging the catenaries accordingly, OAM beams can be produced. Since this geometric phase does not depend on frequency, the~design has a broadband response. {{The~fabrication process of the caternaries is similar to that of the annular apertures in Figure~\ref{continuous}a, except that a 120-nm-thick Au film was deposited instead of the 450-nm-thick silver film.}} The~direction of the grating grooves in Figure~\ref{continuous}c is designed to satisfy $\theta (r,\phi)=l \phi /2$. It is illuminated by a RCP wave and the transmitted LCP wave carries an OAM of order $l$. {{The~gratings were created by etching of a 500-\SI{}{\micro\metre}-thick GaAs wafer through electron cyclotron resonance source (BCl$_3$) to a depth of 2.5-\SI{}{\micro\metre}.}}

\subsection{Metasurfaces for Multiple OAM-Beam Generation}

\begin{figure}[htbp]
	\centering
	\includegraphics[width=5.6in]{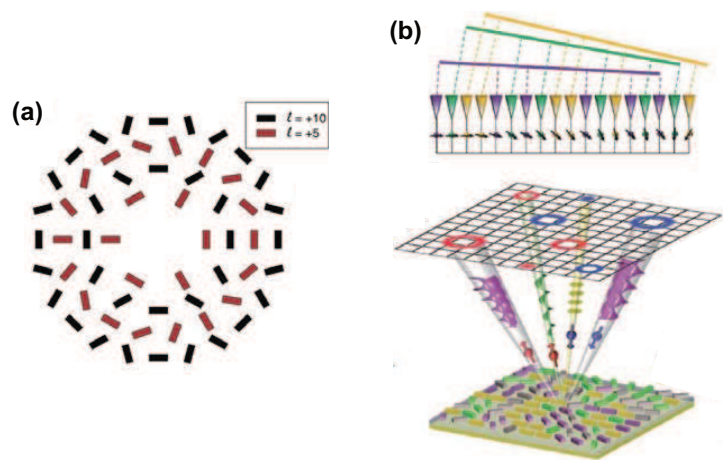}
	\caption{(\textbf{a}) Schematic of the nanofins azimuthal distribution in the inner part of metasurface device with interleaved patterns that generate collinear beams having topological charges $|l|=5$ and $|l|=10$. The~device has a 500 \SI{}{\micro\metre} diameter and contains more than 700 interleaved radial rows of nanofins. Reproduced with permission from~\cite{capasso2017spin}, Copyright Optical Society of America, 2017. (\textbf{b})~Schematic of shared-aperture concepts using interleaved 1D phased arrays and the schematic far-field intensity distribution of wavefronts with positive (red) and negative (blue) helicities. Reproduced with permission from~\cite{hasman2016photonic}, Copyright The~American Association for the Advancement of Science, 2016.}
	\label{multiple}
\end{figure}

Figure~\ref{multiple} depicts two types of geometric-phase metasurfaces for multiple OAM-beam generation~\cite{capasso2017spin,hasman2016photonic}. The~metasurface in Figure~\ref{multiple}a generates two collinear OAM beams. The~dielectric nanofins forming the topological charges of $q=2.5$ and $q=5$ are interleaved each other. {{The~dielectric nanofins consisted of TiO$_2$ and were fabricated based on atomic layer deposition and electron beam lithography.}} In~Figure~\ref{multiple}b, the~nanoantennas with different topological charge are interleaved randomly.

\section{Holographic Metasurfaces for OAM Detection}

\begin{figure}[htbp]
	\centering
	\includegraphics[width=5.6in]{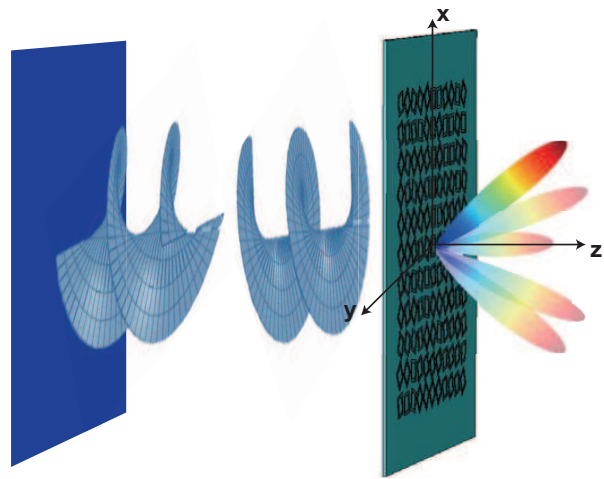}
	\caption{Schematic representation of multiple OAM-beam detection by a single metasurface. Reproduced with permission from~\cite{menglin_detect}, Copyright IEEE, 2017.}
	\label{layout}
\end{figure}

We adopt the holographic concept for the detection of multiple OAM-beam using a single metasurface~\cite{menglin_detect}. The~detection process is summarized in Figure~\ref{layout}. The~metasurface converts the incident wave to multiple waves, only one of which is Gaussian. The~radiation direction of the Gaussian wave is distinguishable according to the order of incident OAM. Consequently, by locating the Gaussian wave, the~incident OAM can be conveniently determined.

The~transmission function of the desired metasurface is given by~\eqref{eq:t}. Then, the~far-field response of the metasurface illuminated by an incident wave carrying OAM of order $l_0$ is calculated by doing the Fourier transform
\begin{equation}
E = F \{E_{in} \cdot t\}=\sum_{m} A_m F \{ E_{\mathrm{OAM}(l_m+l_0)}(k_{xm}, k_{ym})\}.
\end{equation}

Multiple waves with the OAM of order $l_m+l_0$ at the k-space position $(k_{xm},k_{ym})$ are observed. When $l_M+l_0=0$, the~beam is Gaussian and its beam axis is at $(k_{xM},k_{yM})$. It is known that OAM wave has a singularity at its beam axis. Therefore, by examining the field intensity at the positions of $(k_{xm},k_{ym})$, we can identify the gaussian beam, i.e., identify $M$. Then the incident OAM order $l_0$ can be~determined.

\begin{figure}[htbp]
	\centering
	\includegraphics[width=5.6in]{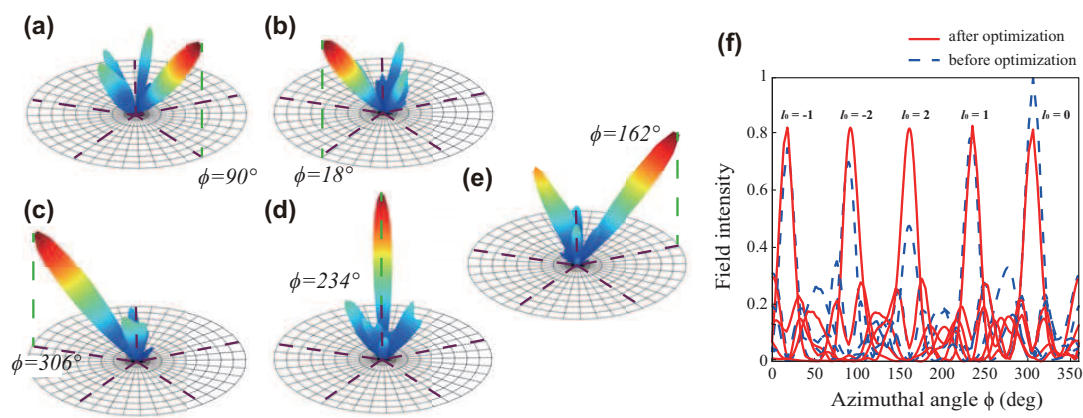}
	\caption{Full-wave simulated far-field power patterns when the incident wave carries OAM of order (\textbf{a}) -2; (\textbf{b}) -1; (\textbf{c}) 0; (\textbf{d}) 1; (\textbf{e}) 2. (\textbf{f}) Original and optimized far-field power patterns at $\theta=40^\circ$ for the five cases (\textbf{a}--\textbf{e}).  Reproduced with permission from~\cite{menglin_detect}, Copyright IEEE, 2017.}
	\label{radpat_cst}
\end{figure}

As a proof of concept, a five-beam case with $l_m=2, 1, 0, -1, -2$ at the directions of $\theta=40^\circ$ and $\phi=90^\circ, 18^\circ, 306^\circ, 234^\circ, 162^\circ$ is demonstrated. The~calculated transmittance $t(r,\phi)$ is implemented using the unit cell as shown in Figure~\ref{unit}a. Full-wave simulated radiation patterns are shown in Figure~\ref{radpat_cst}. It can be seen that the maximum radiation direction (axis of the gaussian beam) when $l_0=-2, -1, 0, 1, 2$ is at $\theta=40^\circ$ and $\phi=90^\circ, 18^\circ, 306^\circ, 234^\circ, 162^\circ$, respectively, which is as expected.

A modified transmission function is proposed to lower the side-lobe level:
\begin{equation}
t_{mod}(r,\phi) = \sum_{m} e^{j(l_m\phi+k_{xm}x+k_{ym}y+\alpha_m)}.
\label{eq_tmod}
\end{equation}

We see an additional phase term $e^{j\alpha_m}$ in~\eqref{eq_tmod}. This~term rotates the $m$th beam and changes the interference status between the beam with other four beams. Therefore, by setting a proper value for $\alpha_m$, we can weaken the constructive interference between adjacent beams, which is the main reason for the high side lobes. The~optimal solution for $\alpha_m$ is $\alpha_1~=~1.0472,\, \alpha_2~=~1.0472,\,\alpha_3~=~2.0944,\, \alpha_4~=~2.7925,\, \alpha_5~=~4.5379$. The~comparison results after optimization are shown in Figure~\ref{radpat_cst}f. When $l_0=\pm1,\pm2$, the~field intensity at a desired location is increased. A general trend of lowered side-lobe level can also be observed. It can be noted that the intensity becomes lower for $l_0=0$. One may break the constructive interference between two adjacent beams but result in an enhanced interference between one of the adjacent beam with the other adjacent beam on the other side. Hence, it is not possible to achieve improvements for all the five incident cases, but there has to be a trade off. Overall, we can observe suppressed side lobes and increased field intensities at desired locations.

\section{Conclusions}

In~summary, we reviewed the research work on the geometric-phase based metasurfaces for OAM generation and detection. The~metasurfaces achieve wavefront manipulation by the spin-induced geometric phase and show high flexibilities. Most importantly, they can be multifunctional. Besides the OAM generation and detection, they are designed for realizing beam multiplexing and demultiplexing and manipulating polarization. Therefore, the~geometric-phase metasurfaces are promising candidates for practical applications of OAM beams.

\acknowledgments{This~work was supported in part by the Research Grants Council of Hong Kong (GRF 716713, GRF 17207114, and GRF 17210815), NSFC 61271158, Hong Kong UGC AoE/P04/08, AOARD FA2386-17-1-0010, Hong Kong ITP/045/14LP, and Hundred Talents Program of Zhejiang University (No. 188020*194231701/208).}
\authorcontributions{Menglin L. N. Chen drafted the manusciprt. Li Jun Jiang and Wei. E. I. Sha revised and finalized the manuscript.}

\conflictsofinterest{The~authors declare no conflict of interest.}

\abbreviations{The~following abbreviations are used in this manuscript:\\

\noindent
\begin{tabular}{@{}ll}
OAM & Orbital angular momentum\\
EM & Electromagnetic\\
AM & Angular momentum\\
SAM & Spin angular momentum\\
LHCP & Left-handed circular polarization\\
RHCP & Right-handed circular polarization\\
LG & Laguerre-Gaussian\\
CGH & Computer generated hologram\\
SPP & Spiral phase plates\\
FSS & Frequency selective surface\\
RCP & Right circularly polarized\\
LCP & Left circularly polarized\\
SRR & Split-ring resonators\\
PEC & Perfect electric conductor\\
PMC & Perfect magnetic conductor\\
PCB & Printed circuit board\\
CSRR & Complementary split-ring resonators\\
LC & Inductor-capacitor
\end{tabular}}




\externalbibliography{yes}
\bibliography{ref_review}


\end{document}